\documentstyle[epsfig,12pt,twoside]{article}
\normalbaselines
\pagebreak

\begin{document}\indent
\hspace{10.35cm}Brown-HET-1210
\vskip 1.5cm
\centerline{\large On Gravity, Holography and the Quantum}
\vskip .75cm
\centerline{\it Carsten van de Bruck$^*$}
\vskip 0.5cm
\centerline{Department of Physics}
\centerline{Brown University}
\centerline{182 Hope Street} 
\centerline{Providence, 02912 Rhode Island, USA}
\begin{abstract}
The holographic principle states that the number of degrees of freedom describing 
the physics inside a volume (including gravity) is bounded by the area of the 
boundary (also called the screen) which encloses this volume. A stronger statement 
is that these (quantum) degrees of freedom live on the boundary and 
describe the physics inside the volume 
completely. The obvious question is, what mechanism is behind the holographic 
principle. Recently, 't Hooft argued that the quantum degrees of freedom on the 
boundary are not fundamental. We argue that this interpretation opens up the 
possibility that the mapping between the theory in the bulk (the holographic theory) 
and the theory on the screen (the dual theory) is always given by a (generalized) 
procedure of stochastic quantization. We show that gravity causes differences to 
the situation in Minkowski/Euclidean spacetime and argue that the fictitious coordinate 
needed in the stochastic quantization procedure can be spatial. 
The diffusion coefficient of the stochastic process 
is in general a function of this coordinate. While a mapping of a bulk theory onto a (quantum) 
boundary theory can be possible, such a mapping does not make sense in spacetimes 
in which the area of the screen is growing with time. This is connected to the average 
process in the formalism of stochastic quantization. We show where the stochastic 
quantization procedure breaks down and argue, in agreement with `t Hooft, that the 
quantum degrees of freedom are not fundamental degrees of freedom. They appear as a 
limit of a more complex process. 
\end{abstract}
\vskip 4.5cm
\centerline{$^*$email: carsten@het.brown.edu}
\vskip 0.1cm
\newpage
\section{Introduction}
The discovery of black hole thermodynamics represents a milestone in the 
search of a consistent unification of the principles of quantum 
mechanics and general relativiy \cite{bek},\cite{bard}. It combines quantum 
mechanics, general relativity and thermodynamics in a unique and fascinating 
picture (for a recent review and discussion, see e.g. \cite{waldi},\cite{jacobson}). 
Yet, the unification between the ideas of general relativity and 
quantum mechanics is not done within a consistent framework, but recent developments in 
string--theory and its relation to M--theory may lead to a self--consistent 
picture. It is interesting to note that some of those developments are sparked also by 
investigations of black holes in the context of these models.

While the origins of the laws of black hole thermodynamics are unknown,
they seem to enforce an upper bound on the number of states within 
a volume \cite{thooft1},\cite{suss}. This so--called 
holographic principle states that the maximal 
entropy within a volume $V$ in space is bounded by its surface area $A$, 
according to the Bekenstein--Hawking formula
\begin{equation}\label{bhentropy}
S_{BH} = \frac{A}{4l_p^2}
\end{equation}
where $l_p$ is the Planck length. It implies that all the physical 
degrees of freedom are somehow encoded on the surface $A$. A somewhat stronger 
statement of this principle would be that a theory which includes gravity, 
describing phenomena within a volume, can be reformulated as a theory 
which describes the evolution of the degrees of freedom on the boundary 
without gravity. 
We refer to this as the weak holographic principle. This has been supported 
at least in some special cases within the framework of the 
Anti--de Sitter(AdS)/CFT--correspondence \cite{malda} and the M(atrix) 
model \cite{matrix}. However, the physical origin of the holographic 
principle remains mysterious. 

In a recent interesting paper, 't Hooft gave a dramatic interpretation of the 
holographic principle (using arguments from black hole physics): 
the fundamental degrees of freedom of nature are not quantum mechanical but 
rather deterministic, in a certain sense classical degrees of freedom \cite{thooft2}. 
Due to information loss these primary degrees of freedom will evolve into a set of 
equivalence classes which (in his definition) evolve unitarily and span a 
Hilbert space. The maximal number of equivalence classes is given by the 
Bekenstein--Hawking formula (\ref{bhentropy}). The information loss mechanism 
was left open in his paper but in the case of classical general relativity 
the information loss could be provided by black holes, as described by 't Hooft. 

If we combine his idea together with the holographic principle in a more general 
framework, we would conclude that
\begin{center}
{\it A classical theory with gravity within a volume $V$ can be formulated 
as a quantum theory with the degrees of freedom living on the boundary 
$\partial V$. The number of quantum states on the boundary is given by (\ref{bhentropy})}. 
\end{center}

We will refer to this principle as the strong holographic principle. According to 
't Hooft the quantum degrees of freedom are {\it not} fundamental.
While ``projecting'' the classical theory onto the boundary (screen), information 
about the classical states would be lost. This, however, has to be 
formulated in a more quantitative way: how does the mapping from the 
bulk onto the boundary theory work? Is this mapping unique or is it different for 
different situations? In any case, spacetime has to know how to dissipate information. 

We should mention that the principle, as formulated above, is supposed to be 
valid in a holographic spacetime, i.e. where the dual theory on the screen exist. 
We will come back to other spacetimes, such as those found in cosmology, in a later section.
There we will give also the argument {\it why} the quantum degrees of freedom are not 
fundamenal. 

Our every-day world is certainly described by quantum mechanics. According to the strong 
holographic principle we could somehow describe quantum degrees of freedom in our 3+1$D$ 
world with a classical theory (including information loss) in a higher--dimensional 
spacetime, whose ``boundary is our 
world''. For example, the Einstein--Podolsky--Rosen correlation between states, 
confirmed by experiments in laboratories in our $3+1D$ world, should be 
classically explainable in a higher--dimensional space--time. Certainly 
gravity may play a fundamental role here as well as the concept of time.

The question we address in this paper is: Can we find a unique way to map the states 
of the classical theory onto the quantum states in the lower--dimensional boundary 
(spacetime)? It would certainly be more satisfying if there would be a unique relationship
between the holographic (bulk) theory and the dual (screen) theory, rather 
than that this relation has to be found seperately for different spacetimes. 
The aim of this paper is to make the first step to find this correspondence 
between the holographic and the dual theory. We assume that this relation 
should be unique (whenever the dual theory exists) and give arguments that 
this is indeed the case.

Our starting point is the observation that there are similar correspondences 
between classical theories and quantum field theories (QFT) even {\it without} 
the inclusion of gravity. These are (we set $c=G=1$):
\begin{itemize}
\item The well--known correspondence between the partition function with 
periodic/ \newline antiperiodic boundary conditions for 
a euclidean quantum field theory in $D$ dimensions and the 
partition function for a classical thermal field theory  
with temperature $ kT = \hbar$, where $k$ is the Boltzmann constant. 
\item Related to this is the stochastic quantization method.
It relates a $D+1$ dimensional classical theory (which includes a stochastic 
noise) to a quantum theory in $D$ dimensions. In the higher--dimensional 
spacetime a stochastic noise plays a fundamental role. 
\end{itemize}
The origin of these relationships is not known, but there 
are strikingly similar to the strong holographic principle. An immediately 
question we ask is therefore, if the strong holographic principle is related 
to these well known correspondences. As we will argue in a later section, 
the correspondence between QFT and classical theories mentioned before is a result 
of the holographic principle in the limit of vanishing gravity. 

The paper is organized as follows: in 
Section 2 we review shortly the concept of stochastic quantization. In 
Section 3 we discuss the strong holographic principle in the limit of 
a flat spacetime. The important case for the Anti--de Sitter (AdS) spacetime 
is discussed in Section 4. In Section 5 we comment on a scalar particle in the 
AdS spacetime and its stochastic quantization. In Section 6 we extend our 
ideas to a spacetime in which the area of the screen is not constant. Our 
conclusions can be found in Section 7, as well as further questions which 
arise in the context of the ideas presented in the paper.

In our discussions we are mainly guided by black hole physics as well as 
expanding spacetimes, such as those which can be found in cosmological 
theories. However, our results are supported by a covariant formulation 
of the holographic principle \cite{bousso},\cite{wald1}. In this paper 
we take a rather heuristic view and formulate the ideas not in a mathematical 
language. We will use the terms ``boundary'' and ``screen'' 
interchangeably.

We mention that other groups are also asking for the mechanism of holography, 
in particular see \cite{sussi},\cite{sussu}.

\section{Review of stochastic quantization}
As mentioned in the introduction, the relation between the holographic 
theory and the dual theory should be unique. It is useful for the discussions
in the later sections to review very briefly the well known relationship 
between classical and quantum field theories mentioned in the introduction. 
In what follows, we mention only the necessary points and refer to the exellent 
reviews \cite{stochast1},\cite{stochast2}.

The starting point is the fact that the Euclidean Green function 
for a scalar field $\phi$ can be interpreted as a correlation function of a 
statistical system in equilibrium of temperature $T = \hbar / k$. 
The Euclidean Green function is\footnote{We consider here the simple example 
for a scalar field. Extensions to more complicated theories such as gauge field
theories exists \cite{stochast1},\cite{stochast2}.}
\begin{equation}\label{euclidean}
\left< \phi(x_1)\phi(x_2)\dots\phi(x_n)\right> =
\frac{\int {\cal D}\phi \exp\left( -(1/\hbar)S_{\rm E}\right) 
\phi(x_1)\phi(x_2)\dots\phi(x_n)}{\int {\cal D}\phi \exp\left( 
-(1/\hbar)S_{\rm E}\right)}.
\end{equation}
Here $S_{\rm E}$ is the Euclidean action. It was the highly ingenious 
idea by Parisi and Wu to interpret the Euclidean path integral measure 
as the stationary distribution of a stochastic process \cite{parisi}. 
This is the basic idea of the procedure known as {\it stochastic 
quantization}. In hindsight it is a rather natural interpretation of 
Feynman path integrals. 

The field $\phi(x)$ in Euclidean space with coordinates $x$ is now 
generalized and will be considered as a function of the Euclidean 
coordinates $x$ {\it and a new fictitious time--coordinate} $t$: 
$\phi(x) \rightarrow \phi(x,t)$. This field couples to a thermal bath.
Let $\eta$ be a Markov stochastic variable, representing the coupling 
of the system to this thermal bath, with temperature $T$ 
\begin{eqnarray}\label{corri}
\left< \eta(x,t) \right> &=& 0; \\
\left< \eta(x_1,t_1) \eta(x_2,t_2) \right> &=& 2 \alpha \delta(x_1 
- x_2)\delta(t_1 - t_2), \nonumber
\end{eqnarray}
where $\alpha$ is the diffusion constant, connected with the temperature $T$ 
(and in general with a friction constant $f$) via
\begin{eqnarray}
\alpha = \frac{kT}{f} \nonumber
\end{eqnarray}
The reason why the new coordinate $t$ is called time is that one imagines 
that the system in $D+1$ dimensions evolves 
in this time. In order that we obtain the usual quantum mechanical 
expressions we have to set $\alpha = \hbar$. 

The basic equation of the stochastic quantization method is the 
Langevin equation
\begin{equation}\label{langevin1}
\frac{\partial \phi}{\partial t} = -\frac{\partial S_E}{\partial \phi} 
+ \eta(x,t),
\end{equation}
where $S_E$ is the action of the field $\phi(x,t)$: 
\begin{equation}
S_E = \int dx {\cal L}(\phi,\partial_x\phi).
\end{equation}
Here, ${\cal L}$ is the Langrangian density. It has the form of the 
original Langrangian, but now one has to replace the field accordingly. 
There is no derivative with respect to the new time--coordinate $t$.
Correlations are now defined as a average over the noise $\eta$.
Then, in this framework, quantum correlation functions in Euclidean 
space are obtained in the limit $t\rightarrow \infty$:
\begin{equation}\label{thermal1}
\left< \dots \right> = \lim_{t \rightarrow \infty} \left< \dots \right>_{\eta}
:= \lim_{t \rightarrow \infty} \frac{\int {\cal D}\eta \exp \left(- \frac{1}{4}
\int dx dt \eta^2(x,t) \right) \dots}{\int {\cal D}\eta \exp \left(- \frac{1}{4}
\int dx dt \eta^2(x,t) \right)},
\end{equation}
where the dots represent solutions of the Langevin equation (\ref{langevin1}). 
One can show that as a result of the evolution of the system within the 
thermal bath the resulting equilibrium distribution is
\begin{equation}
{\cal P}(\phi) \propto \exp\left(-\frac{S_E(\phi)}{\hbar}\right).
\end{equation}
In this equation, the action $S_E$ is evaluated for those $\phi$ which 
satisfy the Langevin equation. 

Equivalently one can find a differential equation, the Fokker--Planck equation, 
for the probability distribution ${\cal P}(\phi)$ of the stochastic process at the 
time $t$. Then quantum Green functions are obtained as
\begin{equation}\label{thermal2}
\left< F(\phi) \right> = \lim_{t\rightarrow \infty} \int
{\cal D}\phi F(\phi) {\cal P}(\phi,t).
\end{equation}
The time evolution for the probability distribution is described by 
the Fokker--Planck equation of the form
\begin{equation}
\frac{\partial}{\partial t}{\cal P}(\phi,t) = 
\frac{d}{d\phi}\left[\frac{d}{d\phi} + \frac{\delta S_E}{\delta \phi} \right] {\cal P}(\phi,t)
\end{equation}
In fact, the stochastic quantization method using the Langevin equation 
is equivalent to the approach starting from the Fokker--Planck equation. 
We refer to the existing literature. 

It should be mentioned that the stochastic quantization method 
can not only be formulated in Eudlidean space but also in Minkowski spacetime. 
Here, the Langevin equation becomes 
\begin{equation}\label{langevin2}
\frac{\partial \phi}{\partial t} = i\frac{\partial S}{\partial \phi} 
+ \eta(x,t),
\end{equation}
where $S$ is the action in Minkowski spacetime. We will, however, mainly 
work in the Euclidean formalism. 

It is clear that the equilibrium limit (\ref{thermal1}) or (\ref{thermal2}) 
must exist in order to make sense for this quantization procedure.
The reason {\it why} the stochastic quantization method works, is yet unknown. 
It is usually taken as a formal manipulation for field theories and it was used 
intensively for computer calculations. In what follows we will argue that 
there is a deeper reason why this procedure works.

\section{The strong holographic principle in the limit of zero gravity}
\subsection{General considerations}
The holographic principle, as originally formulated by 't Hooft, is valid for any 
size and kind of black hole. In what follows we will discuss the case for 
a Schwarzschild black hole. It is well known that the surface gravity $\kappa$ 
for such a black hole is inversely proportional to its mass. Therefore, the larger 
the black hole is, the smaller its surface gravity. To be precise,
\begin{equation}\label{surface}
\kappa = \frac{1}{4M},
\end{equation}
where $M$ is the black hole mass, see e.g. \cite{wald2}.
For a huge black hole the surface gravity is very small, and in the case 
for $M \rightarrow \infty$, $\kappa$ approaches the value zero. But for a 
observer from the outside, all degrees of freedom are still located at 
the horizon of the black hole.

Inside a massive black hole, gravity is less important than it is the 
case for smaller black holes. Take for example the volume ${\cal V}$ within 
which the relation $R_{ijkl}<b$ holds, where $R_{ijkl}\propto M/r^3$ 
is the curvature tensor 
and $b$ is some positive constant. The ratio of the volume ${\cal V}\propto 
r^3$ and the black hole volume ($\propto r_s^3$) will be smaller the larger 
the black hole is:
\begin{equation}
\frac{{\cal V}}{r_s^3} \propto \frac{1}{M^2}.
\end{equation}
On the other hand, let a observer sit at a position $r=v\cdot r_s$, where $v$
is some constant. Then (see e.g. \cite{misner},\cite{joshi})
\begin{equation}
R = {\rm curvature}\mbox{ }{\rm scalar} \propto \frac{M^2}{r^6} \propto \frac{M^2}{v M^6} \propto 
\frac{1}{M^4}
\end{equation}
for consant $v$. Therefore, for an observer, sitting at a constant ratio $r/r_s$, 
gravity will become weaker if the mass of the black hole grows. 
Of course, in the vicinity of the singularity (if there 
is any) gravity is important. However, we neglect for a moment this part 
of the black hole, because the singularity is not a problem for what 
follows\footnote{The asymptotic flatness (see eq. (\ref{surface})) 
is important.}. 
(Although we will come back to the AdS spacetime in the next section, we should 
mention this example, too. The relation between a classical supergravity 
in the AdS spacetime and a quantum supersymmetric Yang--Mills is valid, 
even if there is a black hole embedded in the AdS--spacetime. What is 
important is the asymptotic form of the spacetime, which should be AdS. 
And also, if the radius of the spacetime goes to infinity, 
the space becomes (at least globally) flat, i.e. gravity/curvature goes to zero (for a 
discussion of this limit, see \cite{zerosuss}).)

If the strong holographic principle mentioned in the indroduction makes sense, 
it should be independent of the size of the region, as much as the weak 
holographic principle should be independent of the size of the black hole 
and independent of the size of the AdS spacetime.  
We will argue therefore, that the strong holographic principle is valid 
even in the limit when gravity goes to zero\footnote{It will become clear from 
our discussions that this should be the only way to define the Minskowski--space.}, 
i.e. 
\begin{center}
{\it The strong holographic principle is valid also in asymptotically flat 
spacetimes.}
\end{center}

If we make now the reasonable assumption that the relationship between 
classical theory and the quantum theory is valid independent 
of the spacetime and that the dual theory exists, we postulate that the 
relationship in general has to be given by the stochastic quantization 
procedure. We mention here that the conclusion above is not trivial. The 
strong holographic principle is a relation between a classical theory in 
a volume and a quantum theory on a boundary/screen and is different from 
the original formulation, which states that the number of quantum degrees 
of freedom is bounded by the area of the boundary.

Thus, the thermodynamics of spacetime maybe the origin of the well known 
correspondence mentioned in Section 2. That the lower dimensional theory 
is a quantum theory now follows from the Bekenstein--Hawking formula
(\ref{bhentropy}) with one degree of freedom per Planck unit. Again, 
our conclusion is valid {\bf only} in the case of a spacetime, where 
the limit of the stochastic quantization procedure exist.  
We will return to the implications of other spacetimes below. 
We stress again that the asymptotic behaviour of the spacetime was important 
in the discussion above. 

\subsection{Stochastic quantization in Euclidean/Minkowski spacetime and 
the strong holographic principle}
Let us shortly discuss the stochastic quantization procedure in flat 
Euclidean spacetime in the light of the strong holographic principle. 

As emphasized by Bousso, the screen of Minkowski--space is either 
future or past null infinity when gravity is negligible anywhere. A spacelike 
projection is then allowed as well. The process of stochastic quantization procedure 
is in effect a projection according to the covariant holographic principle. 
The boundary theory should have a ``time--coordinate'', so our intuition suggests 
that the projection should be spacelike. This is indeed possible, because the 
coordinate $t$ in Section 2 has no meaning. Nowhere was it stated that the 
higher--dimensional spacetime should have signature $(D,1)$ or $(D,2)$. The projection 
itself is irreversible when averaged along $t$. 

However, if we take the {\it Euclidean} formulation of stochastic quantization, then 
the situation is as follows. Consider the $D+1$ dimensional Minkowski--space. 
The projection of the theory via stochastic quantization along the $t$--coordinate in this 
space is a dimensional reduction to a $D$ dimensional {\it euclidean} field theory, 
i.e. it gives a euclidean quantum field theory at the screen, which is in this case $I^{+}$. 

\section{The ADS/CFT correspondence and the strong holographic principle}
We have argued that, whenever the dual theory exists, it is related to 
the holographic theory via stochastic quantization. The most impressive 
and explicit example, where a dual theory exists, is the relation 
between {\it classical} supergravity in the AdS spacetime and a 
supersymmetric Yang--Mills theory on the boundary \cite{malda}. More specifically, 
the mathematical formulation of this correspondence is the equivalence 
between the partition functions of both theories \cite{gubser},\cite{witten}:
\begin{equation}
Z_{\rm AdS}(\phi_{\rm bulk}) = Z_{\rm CFT}(\phi_{\rm bound}).
\end{equation}
Here, $\phi_{\rm bulk}$ are the fields in the bulk--theory (a supergravity theory) 
taken at the boundary and $\phi_{\rm bound}$ are the fields in the boundary--theory 
(a supersymmetric Yang-Mills theory). The equation above can be written as\footnote{We 
note here that the original motivation was to relate superstring--theory 
on a AdS$_5 \times S^5$ to a supersymmetric Yang--Mills theory on the 
four--dimensional boundary of the AdS--spacetime. In the equations here 
the partition function of the string theory is approximated as the 
supergravity action. In fact, this is yet the only approximation where a
mathematical formulation of the correspondence exist.}:
\begin{equation}
\exp\left( - \int_{\rm AdS} {\cal L}_{\rm supergravity}(\phi_i(
\phi_i^{\rm bound}))\right)= \left< \exp \int_{\partial {\rm AdS}} 
{\cal O}^i \phi_i^{\rm bound} \right>_{\rm CFT}
\end{equation}
From the point of view of the holographic (bulk) theory, 
the $\phi_i^{\rm bound}$ represent the boundary values of the fields $\phi_i$.
The integral on the left--hand side represent here the classical action 
for the supergravity theory on the AdS spacetime with $d+1$ dimensions evaluated 
at the boundary. On the right--hand side we have the quantum expectation value of the 
primary fields ${\cal O}_i$ of a conformal theory on the boundary, where 
the boundary values $\phi_i^{\rm boundary}$ act as an external source. We 
mention that the radius $R$ of the AdS--spacetime is related to the 
number of colors $N$ and the coupling strength $g_{\rm YM}$ in the Yang--Mills 
theory via $R/l_s = (Ng_{\rm YM}^2)^{1/4}$. 

If our arguments are correct, the supergravity theory in the AdS spacetime and 
the supersymmetric Yang--Mills theory on the boundary should be related by 
stochastic quantization. In the presence of a gravitational field this procedure 
will be modified, as we will discuss in the next section. In the spirit of 
stochastic quantization and with the wisdom of hindsight one dimension 
of the AdS spacetime can be identified with the fictitious time and it is only 
natural to identify this coordinate with the radial coordinate $r$ of the 
AdS--spacetime\footnote{This is agreement with the space--like holographic projection 
described in the work by Bousso \cite{bousso}. A similar remark has been made in the work 
by Lifschytz and Periwal \cite{periwal1}. Their work was connected with the duality between 
string theories and gauge theories and the approach there was different from ours and 
based on the equivalence 
of the Fokker--Planck Hamiltonian for Yang--Mills theories and the loop operator 
\cite{periwal2} (see also the work by Jevicki and Rodrigues \cite{antal}). 
Because the Fokker--Planck equation is at the heart of stochastic quantization, 
Lifschytz and Periwal speculated about the importance of this procedure in the context 
of the AdS/CFT correspondence. We believe that a mathematical rigerous proof 
of our arguments involve indeed the Fokker--Planck Hamiltonian.}. 
In fact, if we insist that during the replacement $\phi(x) \rightarrow \phi(x,r)$ 
the action should be invariant under a supersymmetry, then the spacetime $(x,r)$ has 
to be compatible with this supersymmetry, which in this case is the AdS spacetime. 

In stochastic quantization in Euclidean space with coordinates $x$ and fictitious time $t$ one 
calculates Green functions as an equilibrium limit of the fictitious time, that is 
(see eq.(\ref{thermal1})) 
\begin{equation}\label{stochasticcorrelator}
<P(\phi(x))>=\lim_{t\rightarrow \infty} <P(\phi(t,x))>_\eta,
\end{equation}
where $<...>_\eta$ is the stochastic average and $P$ is some 
polynomial of the fields $\phi$. Now, in the AdS--case the limit 
$r \rightarrow r_{\rm boundary}$, 
correspond to the boundary, where the dual theory lives. Here we see a geometrical 
picture for the stochastic quantization method emerging which is not obvious in the 
case for Euclidean or Minkowski--spacetime (in some coordinate--systems 
$r_{\rm boundary}$ is infinite). In the presence of gravity the fictitious coordinate 
can be a usual spatial coordinate. One may worry that in this case the 
fictitious ``time--coordinate'' is now physically important because fields propagate 
through it and gravity curves it. But because gravity (by holography) localizes the 
quantum degrees of freedom (on the black hole horizon for example) this is what one 
should expect.

What remains to be shown is that $r$ can indeed play the role as the fictitious 
coordinate, that the equilibrium limit exists and that the corresponding theory 
at $r\rightarrow R$ is a quantum supersymmetric Yang--Mills theory. 
What equation (\ref{thermal1}) (with $t=r$) then tells us is actually just 
the statement that correlation functions on the boundary are given by the 
thermal average in the higher--dimensional spacetime with coordinates $r,x$ 
and carrying this to the boundary to the AdS--spacetime. This, by the 
AdS/CFT--correspondence, has to be the correlation function of the 
fields on the boundary\footnote{Here, we can couple the fields onto the boundary. 
We just have to add coupling terms in the Langragian.}. 

Although we have argued that at the heart of the AdS/CFT--correspondence 
is the procedure of stochastic quantization, it is clear that {\it every} theory 
in a AdS spacetime should be related to a quantum theory on the boundary. 
The AdS/CFT correspondence itself is only a special case. We see no reason, {\it why} the 
equilibrium limit for a generalized Langevin equation would {\it not} exist and that 
the whole AdS/CFT--correspondence {\it cannot} be formulated as a problem of 
stochastic quantization. While our discussion was heuristic, we believe that a 
mathematical proof exists. This, however, is beyond the scope of this paper.

\section{A scalar field in AdS}
In this section we give a formal argument that in the case of a scalar field 
our ideas are justified and that the coordinate $r$ can indeed play the role 
of the fictitious coordinate. We show that there is a simple generalization 
of the procedure described in section 2. We are looking now for a stochastic 
process in the higher dimensional spacetime, described by a (generalized) 
Langevin equation. Here we emphazise the physics only, the details of the 
calculations can be found in the appendix.

As discussed in the appendix, we lift the theory into a higher--dimensional 
spacetime and assume the existence of a stochastic process there. The 
arguments there are general and we get a generalized Langevin--equation of the form
\begin{equation}\label{knubbel}
d\phi = -f(r)\frac{\partial S_E}{\partial \phi}dr + dW,
\end{equation}
with 
\begin{eqnarray}
<dW> &=& 0\\
<dW(x,r)dW(x',r')> &=& 2 f(r) \delta(x-x')\delta(r-r') dr.
\end{eqnarray}

In the case of the AdS, the coordinate will now be interpreted as the radial 
coordinate $r$. The metric of this spacetime can be written as 
\cite{suesserwitten}
\begin{equation}
ds^2 = R^2\left[ \frac{4 \eta_{\mu\nu}dx^{\mu}dx^{\nu}}{(1-r^2)^2} 
+ dt^2 \frac{1+r^2}{1-r^2}\right],
\end{equation}
with $r=x_{\mu}x^{\mu}$ and $R$ is the curvature radius of the AdS spacetime. 
As the field is projected along $r$ it is subject to a thermal bath, described 
by the noise--term in the Langevin equation (\ref{knubbel}). It should be noted 
that for this ``fictitious'' process $r$ {\it is} a time--coordinate. 

\begin{figure}
\hspace{3.5cm}\psfig{file=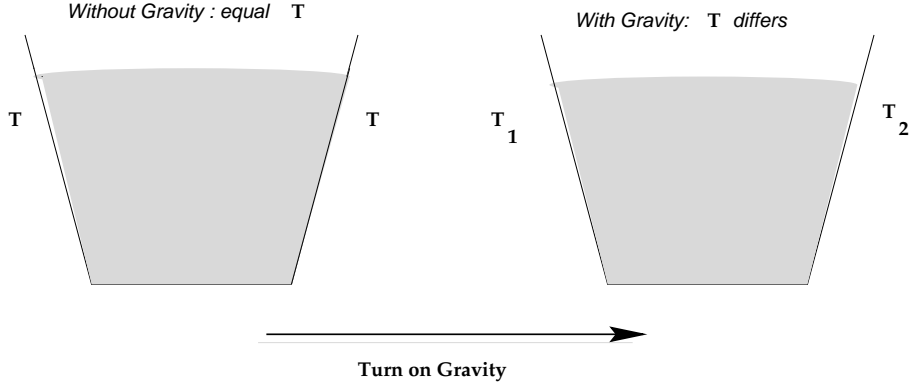,width=12cm}
\caption[h]{Two thermal baths with different boundary conditions. On the 
left--hand side the temperature is the same on both sides. On the right--hand side 
the temperature on one side of the bath is different from the temperature 
on the other side. While the left--hand side represents the situation 
of stochastic quantization in Euclidean/Minkowski--spacetime the right--hand 
side is analogous for the case of a curved spacetime. Gravity causes the 
temperatures (or better: the diffusion parameter) to differ.}
\end{figure}

$f(r)$ is a {\it smooth} function of the radial coordiate $r$ only. 
Here we find a new ingredient in the theory to be discussed: Whereas in 
the ``usual'' stochastical quantization procedure the diffusion constant is 
a real constant, gravity will cause this parameter to be different from point to 
point. A picture might be intuitive (see figure): 
The thermal bath representing the noise $\eta$ has a constant temperature along 
the fictitious coordinate $t$. In the case of the AdS spacetime, and in curved 
spacetimes in general, this temperature will be a function of the fictitious 
coordinate, here the radial coordinate $r$. As discussed in the appendix, the 
fluctuation--dissipation theorem should hold locally.

The probability distribution can be found to be:
\begin{equation}
{\cal P}(\phi) = A \exp(-S_E(\phi(x,r))).
\end{equation}
Here, $\phi$ is a stochastical field for which $\phi(x,r_{\rm boundary}) = \phi(x)$, 
where $\phi(x)$ is the boundary fields. The constant $A$ fixes the normalization 
and therefore we find:
\begin{equation}
{\cal P}(\phi) = \frac{e^{-S_E(\phi)}}{\int {\cal D}\phi e^{-S_E(\phi)}},
\end{equation}
where $\phi$ are solutions of the Langevin--equation (\ref{knubbel}). 
In conclusion, have found that the field on the boundary of the AdS spacetime has the 
usual quantum mechanical expectation values calculated with the Feynman measure for the 
path integrals. 

The example in the Appendix suggests that $f(r)\propto \sqrt{g(r)}$, where $g$ is the 
determinant of the metric of the AdS spacetime, which is a function of $r$ only. More 
importantly, the example suggest further, that the metric (inside the boundary) itself 
is not important. The result (\ref{behav}) can be obtained by stochastic quantization 
in Minkowski space as well as in the AdS case. As pointed out by 't Hooft, the 
holographic principle implies that the geometry inside a volume indeed is unimportant 
\cite{thooft1}. In our approach this is connected to the topological origin of 
stochastic quantization (see the discussion in e.g. \cite{baulieu1}, \cite{baulieu2}, 
\cite{baulieu3} and \cite{yu}). In fact, it was shown that the stochastically quantized 
theory is equivalent to a topological field theory. Furthermore, we recall here the 
well known fact, that, at least in some cases, a classical stochastic process can be 
reformulated as supersymmetric quantum mechanical problem, see e.g. \cite{parisi2} and 
\cite{junker} and references therein.

To complete our discussion, we have to consider the entropy bound (\ref{bhentropy}) 
in the framework of our theory. Unfortunately we were not able to find a way how 
our ideas lead to a {\it derivation} of the entropy bound. What we can only say is that 
the theory on the boundary has to be a quantum theory. Because of the covariant entropy 
conjecture formulated by Bousso \cite{bousso} and the modified version by Flanagan, Marolf 
and Wald \cite{wald1} the bound should be satisfied. All what has to be assumed is 
that the conditions for the holographic projection has to be fulfilled. The projection 
via stochastic quantization in the case of the AdS spacetime disussed here is a spacelike 
projection in the sense of Bousso. It is interesting to speculate that the saturation of the 
entropy (see eq. (\ref{bhentropy})) at the boundary is connected to the fact that the 
stochastic process reaches an equilibrium state where the entropy simply does not 
grow.

A crucial point in the discussion so far is that the ``temperature'' (or better: diffusion 
coefficient) along the boundary should be independent of the euclidean time, otherwise a 
sensible limit would not exist. This is important for what follows.

\section{Holography in cosmological spacetimes: implications}
Our discussion so far was based on the case of a spacetime 
where the holographic theory and the dual theory exist. Certainly our 
universe is dynamical and expanding. 

That the weak holographic principle has to be modified in more 
general spacetimes, such as those in cosmology for example, was 
discussed by Fischler and Susskind \cite{fischler}, 
Bak and Rey \cite{Bak}, Veneziano \cite{veneziano}, 
Easther and Lowe \cite{easther}, and Kaloper and Linde \cite{kaloper}, see 
also the discussions in 
\cite{rama},\cite{burstein},\cite{venezia},\cite{brusi},\cite{ellis}.
Based on these earlier ideas, Bousso formulated a general covariant 
holographic principle \cite{bousso}. In what follows we will not be able to 
give a general theory. Rather we will discuss the implication for the suggestion 
by 't Hooft, that the fundamental degrees of freedom are not quantum mechanical.

The duality between quantum and classical theories, as stated by the 
strong holographic principle, answers the question which theory is 
``more fundamental'', because at this point both theories are not on the same 
footing. We consider several gedankenexperiments in this section which makes 
this point clear.

Suppose that our universe was in its earlier epoch in a AdS--spacetime state and
(approximately) static. All degrees of freedom live on the boundary of the 
AdS--spacetime and all processes in the bulk can be described by the boundary theory. 
Let there now be a process which turns the state of the universe into another one, 
say a matter dominated phase. This can happen for example if the cosmological 
constant becomes positive through some dynamical processes and decays into 
particles. What happens to the boundary 
in this process? An observer in the bulk will be able to see only a part of 
the de Sitter--space. The horizon is $H^{-1}$, where $H$ is the expansion rate. 
This horizon gets dynamical when the universe becomes matter dominated. 
In such cases it was proposed that the holographic principle should be replaced 
by the (generalized) second law \cite{easther}. This would imply that it is not longer useful 
to talk about a quantum boundary theory, because degrees of freedom maybe created or 
destroyed. It is difficult to see if such a theory is compatible with 
the second law as well as with unitarity in general; in short: there will be no 
quantum mechanical degrees of freedom. 

What happens to the bulk theory? It would hardly makes sense if the theory, 
which is classical in the beginning becomes now ``more and more'' quantum. 
It is more likely that the theory remains classical. Of course, our discussion 
implies that the theory on the boundary can only exist if the holographic 
theory exist, because the spacetime in this theory was the starting point. 
This implies, however, that the quantum degrees of freedom are not the fundamental 
ones but the classical degrees of freedom in the bulk. This is in agreement with 
't Hooft's proposal (and the philosophy of stochastic quantization itself). 

Another example would be a black hole. We call all degrees of freedom on the boundary
quantum mechanical when the black hole surface is static and non--growing, but we just 
have to throw matter in the black hole to destroy the quantum character of the degrees 
of freedom. During this process, the total entropy will grow. 

Given the ideas in this paper, the fundamental difference 
between a spacetime with constant screen area and a general expanding spacetime 
seems to be that the former allows for a limit of the stochastic process, i.e. 
the stochastic process described by the variable $\eta$ is in local thermodynamical 
equilibrium and the procedure of stochastic quantization makes sense, i.e. the 
limit $t\rightarrow \infty$ exists (see eq. (\ref{thermal1})). As explained 
in the last section, the temperature for $T(r\rightarrow r_{\rm boundary(1)})$ and the 
temperature $T(r\rightarrow r_{\rm boundary(2)})$, where the fictitious coordinate $r$ runs 
from $r_{\rm boundary(1)}$ to $r_{\rm boundary(2)}$, must be constant. 
However, there is no reason to believe that 
this is always the case. Non--equilibrium processes in the holographic theory 
are related to the fact, that, to use the words of 't Hooft, the equivalence classes, 
which may form, do not evolve not unitarily in this case and the corresponding 
quantum theory does not exist. In this context it is 
interessting to note that Marchesini has shown that the loop equations of 
non--abelian gauge theories are equivalent to the equilibrium condition 
within the context of stochastic quantization \cite{march}. 
In fact, stochastic quantization only 
makes sense (and is defined in that way) as an equilibrium limit 
(for $t \rightarrow \infty$). In conclusion: a dual theory always exists 
when a sensible equilibrium limit for $t \rightarrow \infty$ exists, because
then stochastic quantization is applicable, and the condition for that is 
connected to the spacetime structure, described by the holographic theory. 
The screen, on which the degrees of freedom are projected, has to allow 
for a sensible limit\footnote{The screen in AdS for example allows for such a 
projection.}. We see here a connection of our argumentation to 
the work by Easther and Lowe, who in particular argued that in the case of  
general spacetimes the holographic principle has to be replaced by the second 
law: if the entropy of a system grows, it is not in thermodynamical equilibrium 
and hence a stochastic quantization procedure makes no sense. The black hole 
example mentioned above is a good example for what is going on here. 

It is interesting to speculate on the role of supersymmetry. Supersymmetry 
is connected to stochastic processes in a subtle way \cite{parisi2}, because 
of the hidden supersymmetry in the Langevin and Fokker--Planck equation. 
It is well known that supersymmetry is possible only in certain spacetimes, 
such as AdS, but not, for example, in an expanding universe. But here, stochastic 
quantization breaks down, too. 

We can draw an important conclusion from the discussion above:
{\it In some very strong time--varying gravitational fields, where the 
holographic principle makes no sense and has to be replaced with the 
generalized second law, we expect significant deviations from the quantum 
mechanical predictions.} This is because the relationship between 
quantum--mechanical and classical theories is lost in those strong 
time--variating gravitational fields, similar to the process of dropping matter 
into a black hole. There the degrees of freedom might be destroyed 
or generated, which differs from ordinary quantum mechanical behaviour. One possible 
observational consequence would be the violation of unitarity in our $3+1D$ world 
in such strong time--varying gravitational fields.

\section{Conclusions and outlook}
The strong holographic principle mentioned in the introduction is stronlgy 
connected to the well known relationships between classical field 
theories, statistical mechanics and quantum theory in Minkowski/Euclidean space. 
In this paper we gave some arguments why this is the case. We have argued 
that whenever the dual boundary theory exists, the relationship between the 
holographic (bulk) theory and the dual (screen) theory should be unique, i.e. 
independent of the spacetime, so that the strong holographic principle is valid 
even in the limit of vanishing gravity (curvature). Because the Minkowski spacetime 
can (and should) be viewed as a limit process of vanishing gravity, we argued that 
one can here also find a relationship between classical and quantum mechanics. The 
natural candidate we propose is the stochastic quantization procedure, which relates a 
$D+1$--dimensional stochastic classical theory to a $D$--dimensional quantum theory. 
This relationship should hold whenever a dual theory exist, especially in the case 
of the AdS/CFT correspondence. We have argued (but not shown), that the 
AdS/CFT--correspondence can indeed be seen as a process of stochastic quantization, 
where this process has a geometrical meaning in the AdS--spacetime. We have argued, 
that the radial coordinate of the AdS spacetime can act as the stochastic time. 
Furthermore, we argued that stochastic quantization has the property that the (smooth) 
interior metric has no effect on the physics on the boundary. This property of holography 
was first discussed by 't Hooft \cite{thooft1}.

The idea by 't Hooft of the emergence of quantum degrees of freedom was based on 
information loss at the classical level. While it has to be shown how exactly quantum 
states emerge from classical states, the information loss was here provided by the 
noise. What is not known at this point is the origin and the physical meaning 
of that noise, which has to be included for the stochastic quantization 
procedure\footnote{In \cite{mueller} it was argued that chaos may play a significant role. 
The authors mention the important example 
of classical Yang--Mills theories, which are chaotic dynamical systems \cite{mueller2}. 
In their work the Langevin--equation plays an essential role, which provides a 
effective description on large scales. For another 
approach on holography and chaos see \cite{arefa}.}. We believe that the noise 
is not an artifical quantity in the sense that it has no physical meaning. Rather, 
it seems to be connected with the coupling 
of matter and spacetime itself. We note here that we don't believe that 
the Langevin--equation is a fundamental description, but an effective one. 

In a sense, if the ideas presented here have something 
to do with reality, the existence of quantum degrees of freedom which we observe in the 
laboratory are a result of the existence of extra dimensions. 
Our arguments suggest strongly that quantum mechanical expectation values should be 
seen as an average over a stochastic process and are therefore not fundamental quantities 
itself. This stochastic process seems to be connected to spacetime in a subtle way. Obviously, 
there is a connection with the interpretation of quantum mechanics by Nelson \cite{nelson}, 
and that spacetime itself is the source for the stochastic noise needed in this work.

A lot of work remains to be done. Most importantly, we left open if the 
AdS/CFT--correspondence can really be understood as a process of stochastic quantization. 
It is very important to investigate this case, first because it can confirm (or 
not confirm) our ideas. Secondly, if it turns out that the stochastic quantization 
is at the heart of holography ({\it in every spacetime}), the AdS spacetime is a 
very good example where one can learn more about the thermodynamics/statistics 
of this spacetime. Furthermore, one may hope to find hints about the nature of 
the stochastic noise. The way pioniered by Periwal and Lifshytz should tell us more. 
Our approach should also be discussed within the framework of M(atrix) theory: whereas we 
worked in the spacetime picture, the ideas presented here should have a 
more fundamental interpretation. 

Apart from the case of the AdS spacetime, one should consider the case of a black hole 
in the light of the ideas presented here. Of course, a covariant formulation of the ideas 
presented here would be desireable. 

Another important problem to be solved is to find a mathematical expression for the 
condition that the dual quantum theory exists. We believe that this question is deeply 
related to the thermodynamics of spacetime and therefore to the generalized second law. 

\begin{center}
{\bf Acknowledgements:}
\end{center}
We thank Stephon Alexander, Robert Brandenberger, Miquel Dorca, Damien Easson, 
Antal Jevicki, Jerome Martin, Matthias Soika and Shan-Wen Tsai for useful 
discussions and critism at several stages of this project and for making the 
paper readable. We are grateful to Helmuth H\"uffel for pointing out some 
useful references and comments. This work was supported by DAAD/NATO.

\begin{appendix}
\section{Treatment of the Langevin equation}
In this appendix we justify our steps in Section 5. The field $\phi(x)$ in 
Euclidean space (with coordinates $x$) is now ``lifted'' onto a 
higher--dimensional manifold with an additional coordinate $r$ and metric 
$g_{\mu\nu}$, which will be here the AdS--spacetime. On this manifold we postulate a 
stochastic process $\eta$ and imagine, that the generalized scalar field 
$\phi(x,r)$ is ``propagating'' along the coordinate $r$. We set $\hbar=c=G=k=1$.

Usually the AdS is considered as a submanifold in a higher--dimensional 
covering space with symmetry group $SO(2,D)$. However, we make use 
of the (euclidean) form (see e.g. \cite{suesserwitten})
\begin{equation}\label{ads}
ds^2 = R^2\left[ \frac{4}{(1-r^2)^2}\left( dr^2 + r^2 d\Omega^2\right) 
+ d\tau^2 \frac{1+r^2}{1-r^2}\right]
\end{equation}
where $r<1$ is the AdS spacetime and $r=1$ is the boundary.

Before we discuss the specific example of the AdS spacetime, let us make some 
general considerations. Consider a point $(x,r)$ on the higher--dimensional 
spacetime. The boundary of this spacetime is our original euclidean space. 
We postulate a stochastical differential equation which may viewed as a 
``generalized'' Langevin--equation\footnote{This is 
actually not the Langevin--equation as in the usual stochastic quantization 
procedure because the cofficients depend on the coordinate $r$. The equation 
has the form of what is called It$\hat{o}$'s Langevin equation. In what 
follows, the eq. (\ref{sabina}) only has to allow for the correct limit, 
i.e. we want to recover the euclidean measure in the path integral as $r$ approaches 
the boundary.}. Because we are in a curved space we have to treat the problem locally, 
that is, we write generally
\begin{equation}\label{sabina}
d\phi = -\Gamma(x,r) \frac{\partial S_E}{\partial \phi(x,r)} dr + dW(x,r),
\end{equation}
where we write formally ``$dW(x,r) = \eta(x,r)dr$''. The meaning of $\Gamma$ will become 
clear if one notices that this quantity ``absorbes'' the change of a coordinate transformation 
$r \rightarrow \tilde{r}$. It describes also the strength of dissipation in the Langevin 
equation. Because $\phi(x,r)$ is a scalar function, as well as the Euclidean action, 
$dW(x,r)$ has to be a scalar function, too. For the transformation for $\Gamma$ we 
find therefore
\begin{equation}
\Gamma' = \Gamma\frac{dr}{d\tilde{r}}.
\end{equation}
In what follows, we will consider $\Gamma$ and $dW$ as a function of $r$ only. This should 
be the case in the AdS, for example, reflecting the symmetries of this space\footnote{In 
fact, the symmetries of the AdS spacetime motivated us to choose the form (\ref{sabina}) 
for the Langevin--equation and the form of the correlations below.}. 
The correlation for the process $dW$ is assumed to be
\begin{displaymath}
<dW>=0 \mbox{ }\mbox{ }{\rm and} \mbox{ } <dW(r)dW(r')>
=2 \alpha(r)\delta(r-r') dr,
\end{displaymath} 
$\alpha(r)$ describes the strength of the fluctuations and is therefore related to 
the local temperature of the bath. The transformation of 
$\alpha(r)$ is:
\begin{equation}
\alpha' = \alpha \frac{dr}{d\tilde{r}}.
\end{equation}
$\alpha$, the diffusion coefficient, is therefore in general not 
constant along the coordinate $r$.\footnote{We may introduce another stochastic variable which has a 
constant temperature along the radial coordinate. This variable transforms then as the 
variable $\Gamma$.} Finally we have to specify our boundary conditions for the Langevin--equation, 
which is
\begin{equation}
\lim_{r\rightarrow r_{\rm boundary}} \phi(x,r) = \phi(x),
\end{equation}
i.e. the field is the original field when we approach the boundary 
(which could also be the event horizon in the case of a black hole). Of course, it is 
clear that the generalized temperature should be a smooth function as we approach 
the horizon/boundary. Furthermore, we assumed that the notion of temperature locally makes 
sense, as usual in non--equilibrium thermodynamics. However, what is the 
Langevin equation for this problem? Are there conditions for $\alpha(r)$ and 
$\Gamma(r)$? And can we find the solution we want? 

In order to answer these questions, we will shall now derive the Fokker--Planck equation for this 
problem, using Ito's stochastic calculus. Consider a functional $F(\phi(x,r))$ of $\phi(x,r)$. The 
Taylor series is, using the Langevin equation
\begin{eqnarray}
dF(\phi) &=&  \frac{\delta F}{\delta \phi} d\phi 
              + \frac{1}{2}\frac{\delta^2 F}{\delta \phi^2} d\phi^2+... \nonumber \\
         &=&  - \Gamma \frac{\delta F}{\delta \phi}\frac{\delta S}{\delta \phi} dr 
              + \frac{\delta F}{\delta \phi}dW + \frac{1}{2}\frac{\delta^2 F}{\delta \phi^2}
                \left[ \Gamma^2 \left( \frac{\delta S}{\delta \phi} \right)^2 dr^2 + ... + dW^2 \right]
              + ...
\end{eqnarray}
If we neglect the higher order terms and take the average of this equation we find
\begin{equation}\label{eins}
\frac{d<F(\phi)>}{dr} 
= -\left< \Gamma\frac{\delta F}{\delta \phi}\frac{\delta S}{\delta \phi} \right> 
  + \left< \alpha\frac{\delta^2 F}{\delta \phi^2} \right>,
\end{equation}
where we used the conditions for the stochastic noise. Now we introduce a probability 
distribution ${\cal P}(\phi,r)$, defined by
\begin{equation}
<...> = \int {\cal D}\phi {\cal P}(\phi,r) ...,
\end{equation}
where the dots represent a polynom in the field $\phi$. It is 
\begin{equation}
\frac{d<F(\phi)>}{dr} = \int {\cal D}\phi \frac{\partial {\cal P}}{\partial r} F(\phi).
\end{equation}
Integrating (\ref{eins}) by parts we then find the Fokker--Planck equation
\begin{equation}
\frac{\partial {\cal P}}{\partial r} = \Gamma(r) \frac{\delta}{\delta \phi}
\left( \frac{\delta S}{\delta \phi} {\cal P} \right) 
+ \alpha(r)\frac{\delta^2}{\delta \phi^2} {\cal P}.
\end{equation}
We will now investigate if we can find a solution of the form
\begin{equation}
{\cal P}(\phi,r) = A(r)e^{-S_E (\phi(x,r))}.
\end{equation}
Inserting this into the Fokker--Planck equation gives
\begin{equation}\label{lab}
\frac{A'(r)}{A(r)} = \left(\Gamma(r)-\alpha(r)\right)
\left( \frac{\delta^2 S}{\delta \phi^2} - \left(\frac{\delta S}{\delta \phi}\right)^2\right).
\end{equation}
From this equation we find with find, using seperation of variables, 
for $A(r)$: 
\begin{equation}
A(r) = {\cal C} \exp \left( {\cal B} \int_{0}^{r} d\tilde{r} \left(\Gamma(\tilde{r}) 
- \alpha(\tilde{r}) \right) \right),
\end{equation} 
and for $\phi$:
\begin{equation}
\frac{\delta^2 S}{\delta \phi^2} - \left(\frac{\delta S}{\delta \phi}\right)^2 = {\cal B}.
\end{equation}
The last equation is not consistent to solve because $S_E$ is a free function 
of $\phi$. If we use $\Gamma(r)=\alpha(r)$ in the Langevin--equation (\ref{sabina}) 
from the very beginning of the calculation, we would find from equation (\ref{lab}) 
that $A(r)={\cal C}=$ constant and no restriction to $S_E$ would apply.
The condition $\Gamma(r)=\alpha(r)$ is a result of the well known fluctuation--dissipation 
theorem which should hold at every point. 
We fix ${\cal C}$ by the requirement
\begin{equation}
\int {\cal D}\phi {\cal P}(\phi) = 1.
\end{equation}

In conclusion, we can formulate the problem as a stochastic process with the Langevin 
equation is given by eq. (\ref{sabina}) if we assume the fluctuation--dissipation theorem 
(here in its local form). Furthermore, the temperature along the fictitious coordinate is 
not constant. The probability distribution is given by 
\begin{equation}
{\cal P}(r,\phi) = \frac{e^{-S_E(\phi(x,r))}}{\int {\cal D}\phi e^{-S_E(x,r)}}.
\end{equation}
Because $\phi(x,r) \rightarrow \phi(x)$ as we reach the boundary of the spacetime, 
the probability distribution approaches the euclidean path integral measure for the 
field living there\footnote{We remember again that $S_E$ is the euclidean action for the 
boundary field $\phi(x)$.}.

As a last step we have to find an expression for the function $\Gamma$. 
It is instructive to discuss a simple example in zero dimensions. Although this example 
is simple, it will illuminate two important points: first we will find an expression for 
$\Gamma$ and second we will argue that the metric of the higher dimensional space is 
unimportant. 
 
Consider the field $\phi(r)$ with the action
\begin{equation}
S_E = \frac{1}{2}m^2\phi^2.
\end{equation}
We want to calculate the correlation function $<\phi^2>$. With eq. (\ref{eins}) it 
follows that 
\begin{equation}
d<\phi^2> = -2\Gamma(r) m^2 <\phi^2> dr + 2\alpha(r) dr.
\end{equation}
For the solution we make the ansatz ($C$ is constant):
\begin{equation}
<\phi^2> = Ce^{- \beta(r)} + A(r).
\end{equation}
Then we find the following conditions
\begin{eqnarray}
2\Gamma(r)m^2 &=& \beta'(r), \\
2\alpha(r) &=& A'(r) + \beta'(r)A(r).
\end{eqnarray}
According to the fluctuation--dissipation theorem $\Gamma(r)=\alpha(r)$. Then a solution is 
\begin{equation}
<\phi^2> = C \exp \left(-2 m^2  \int_{0}^{r} \Gamma(\tilde{r})d\tilde{r}\right) + \frac{1}{m^2}.
\end{equation}
Because $\Gamma(r)$ is a positive function we find for $r\rightarrow \infty$ 
(which may correspond to the boundary in a certain coordinate system) 
\begin{equation}\label{behav}
<\phi^2> \rightarrow  \frac{1}{m^2},
\end{equation}
which can also be obtained using the path integral with the measure $\exp(-S_E)$. 

One might wonder that in the calculation above the explicit form of $g_{\mu\nu}$ 
was not used. In fact, the only ingredient was that the diffusion parameter of the fictitious 
bath depends {\it only on the fictitious coordinate $r$}, i.e. the radial coordinate 
of the AdS. $\Gamma(r)$ must have a singular behaviour there in order that eq. (\ref{behav}) holds.
Now we observe that the transformation of $\Gamma(r)$ is the same as 
a usual tensor density, such as the determinant of the metric tensor. This suggests, that we 
could make the ansatz ($p$ is a positive constant)
\begin{equation}\label{ansatz}
\Gamma(r) = p \sqrt{-g},
\end{equation}
because in the case of AdS the determinant of the metric tensor {\it is a function of }
$r$ {\it only}: $g=g(r)$. It is singular at the boundary. Indeed, with this ansatz we 
find in the coordinates (\ref{ads}) (reducing to one dimension) the behaviour (\ref{behav}) 
for the correlation function. Furthermore, the result (\ref{behav}) can be obtained if one 
considers a Minkowski--space instead of an AdS space (see e.g. \cite{stochast2}) . This point 
is disussed further in Section 5. 
\end{appendix}

\end{document}